\documentclass[aps,prb,reprint,twocolumn,superscriptaddress,longbibliography]{revtex4-2}
\pdfoutput=1

\usepackage{graphicx}
\usepackage{amsmath,amsfonts,amssymb}
\usepackage{bbold}
\usepackage{physics}
\usepackage{color}
\usepackage{hyperref}
\usepackage{array}
\usepackage{tabularx}
\usepackage{booktabs}
\usepackage{makecell}
\usepackage{textcomp}

\definecolor{orange}{rgb}{1.0, 0.5, 0.0}

\DeclareMathOperator{\diag}{diag}

\begin{document}

\title{Local signatures of altermagnetism}

\author{Jannik Gondolf}
\affiliation{Niels Bohr Institute, University of Copenhagen, DK-2200 Copenhagen, Denmark}

\author{Andreas Kreisel}
\affiliation{Niels Bohr Institute, University of Copenhagen, DK-2200 Copenhagen, Denmark}

\author{Mercè Roig}
\affiliation{Niels Bohr Institute, University of Copenhagen, DK-2200 Copenhagen, Denmark}
\affiliation{Department of Physics, University of Wisconsin–Milwaukee, Milwaukee, Wisconsin 53201, USA} 

\author{Yue Yu}
\affiliation{Department of Physics, University of Wisconsin–Milwaukee, Milwaukee, Wisconsin 53201, USA} 

\author{Daniel F. Agterberg}
\affiliation{Department of Physics, University of Wisconsin–Milwaukee, Milwaukee, Wisconsin 53201, USA} 

\author{Brian M. Andersen}
\affiliation{Niels Bohr Institute, University of Copenhagen, DK-2200 Copenhagen, Denmark}

\date{\today}

\begin{abstract}
Altermagnets constitute a class of collinear compensated Néel ordered magnets that break time-reversal symmetry and feature spin-split band structures. Based on versatile microscopic models able to capture the altermagnetic sublattice degrees of freedom, we study characteristic local signatures of altermagnetism near disorder sites. We give a complete list of two-dimensional models that exhibit altermagnetism classified by their corresponding layer groups. Specifically, we calculate the local density of states in the vicinity of pointlike nonmagnetic impurities and expose its spatial dependence for two minimal models showcasing $d$-wave and $g$-wave altermagnetism. The momentum structure of the nodes ($d$-wave, $g$-wave, etc.) is directly imprinted on the total local density of states, thus measurable by scanning tunneling conductance experiments. This signature is present both in the spin-resolved as well as the spin-summed local density of states. We find a weaker response in the nonmagnetic state from the anisotropic crystal environment and uncover the importance of the sublattice degree of freedom to model altermagnets. We also study coexistence phases of altermagnetism and superconductivity and provide predictions for the local impurity response of in-gap bound states. The response of impurity bound states strongly enhances the distinct altermagnetic signature.
\end{abstract}

\maketitle

\section{Introduction}

Altermagnets have been identified as a distinct class of collinear compensated magnetic order in addition to standard antiferromagnetism~\cite{Smejkal_Emerging_2022,Smejkal_Conventional_2022}. As opposed to conventional antiferromagnets, Néel ordered altermagnets lack inversion or translation together with time-reversal as a combined symmetry, but exhibit instead time-reversal and a rotation (proper or improper) as a combined symmetry of the magnetic state.
This distinction can originate from anisotropic local crystal environments around the different sublattice sites. Thus, the Néel order of altermagnets exhibits the same unit cell as the original normal state. The unique properties of altermagnets have profound implications for their electronic properties, most notably the existence of large momentum-dependent spin-split electronic bands, even in the absence of relativistic spin-orbit coupling. Spin splitting has recently been detected by angular-resolved photoemission spectroscopy (ARPES) on a number of altermagnetic candidate materials, for example CrSb~\cite{Reimers2024,Ding2024}, MnTe~\cite{Krempasky_Altermagnetic_2024,LeeMnTe2024,Osumi2024,Osumi2024}, and RuO$_2$~\cite{Fedchenko}.
The altermagnetic spin-split bands that arise in a material with negligible net magnetization and absence of harmful stray fields have led to the proposal of several applications of altermagnetism within spintronics~\cite{Smejkal_Emerging_2022}.
However, a fundamental challenge is the presence of domains that may limit their applicability and complicate the experimental interpretation of the unique signatures of altermagnets. The latter motivates studies of local signatures of altermagnetic order. What should local atomic-scale resolved probes detect as characteristic features of altermagnetic order? From studies of the local impurity response in unconventional superconductors, for example, it is well known that impurities can provide useful information on the fundamental properties of the host~\cite{Pan2000,Hudson2001,Balatsky_RMP,Kreisel2015,Shun,Choubey2017}.

Here we address the question of local signatures of altermagnetic order, starting from recently developed microscopic minimal Hamiltonian models for altermagnets. The framework developed in Ref.~\cite{Roig_Minimal_2024} provided 40 electronic models for all centrosymmetric space groups with magnetic atoms occupying
inversion-symmetric Wyckoff positions of multiplicity two. This includes monoclinic, orthorhombic, tetragonal, rhombohedral, hexagonal, and cubic materials and describes $d$-wave, $g$-wave, and $i$-wave altermagnetism~\cite{Roig_Minimal_2024}. Thus, the minimal models are general, yet may also become material-specific through parameter constraints from Density Functional Theory (DFT) or experiments and respect all symmetries of the crystal in the normal state and the altermagnetic state.
Clearly, a significant advantage of having such minimal models is the ability to perform analytical calculations and derive general formulas that apply to broad classes of altermagnetic materials. Examples of this approach are found in Refs.~\cite{Roig_Minimal_2024,Roig_Landau} where general expressions for the Berry curvature are reported, allowing for calculations of the anomalous Hall effect (AHE), resulting in the insight that altermagnets exhibit AHE linear in their spin-orbit coupling. Other recent examples of the usefulness of microscopic minimal models include derivations of connections between quantum geometry and altermagnetic instabilities~\cite{Heinsdorf}, and studies of altermagnetic anti-chiral surface states and domain wall bound states~\cite{Sorn}.

In this work, we apply microscopic minimal Hamiltonian models for altermagnetism to obtain the local Green's function near disorder sites. For concreteness and simplicity, we focus on minimal models that can be obtained from three-dimensional (3D) tight-binding models as presented in Ref.~\cite{Roig_Minimal_2024} by removing the $k_z$ dependence. These models belong to space groups 11, 13, 14, 51, 55, 83, 123, and 127 and have either $d$-wave or $g$-wave spin-split bands. In Appendix~\ref{Yuetable}, we give the corresponding 2D models as classified by the corresponding 2D layer groups.
As deduced from Tab.~\ref{tab:layerGroups}, the models considered in this work are relevant for layer groups L17, L44, L63 and, in a special case, to L51.
For a high-throughput computational search of 2D altermagnetic candidate materials, we refer to Ref.~\cite{ThomasOlsen} which includes a series of 2D candidates for which our models apply. We stress, however, that the results of local signatures of altermagnetic order obtained below may also apply to 3D altermagnets, depending on their particular surface termination. In order to address the impurity problem, we apply both a $T$-matrix formulation and perform real-space self-consistent calculations to determine the local density of states (LDOS) response near point-like impurities. We find that the altermagnetic characteristics are directly imprinted on the total LDOS, allowing for detection of local altermagnetism and its associated momentum structure. Finally, we also discuss the impurity properties of altermagnets coexisting with superconductivity. In this case, in-gap impurity bound states are generated by both magnetic and nonmagnetic pointlike impurities. The bound state wavefunction feature clear signatures of the underlying altermagnetic order.

Some earlier work has addressed local markers for altermagnetic order, including bound states near lattice dislocations~\cite{Zhu2024}, spin-polarized subgap states in altermagnetic superconductors~\cite{Maiani}, and characteristics of impurity-induced Friedel oscillations in altermagnetic metals~\cite{Sukhachov,Hu_Quasiparticle_2025,Chen_impurity2024}. Here, we focus on the role of pointlike nonmagnetic impurities. As shown below, our results are qualitatively different from those presented in Refs.~\cite{Maiani,Sukhachov,Hu_Quasiparticle_2025,Chen_impurity2024}. The important distinction originates from the applied minimal model used to describe altermagnetic materials. Whereas the results in Refs.~\cite{Maiani,Sukhachov,Hu_Quasiparticle_2025,Chen_impurity2024} were based on a one-band model with momentum-dependent spin-splitting, our minimal models for altermagnetism are based on fundamental symmetries, i.\,e. the relationship between the site symmetry of magnetic atoms and the point group symmetry of the space group~\cite{Roig_Minimal_2024}. As a consequence, such minimal models incorporate the (lower) symmetry already in the nonmagnetic normal state. Indeed, in the minimal models we consider here, the instability into the altermagnetic state is driven by antiferromagnetism and the altermagnetic spin-spitting is an induced order parameter. In single band models of Refs.~\cite{Maiani,Sukhachov,Hu_Quasiparticle_2025,Chen_impurity2024}, the altermagnetic spin-splitting is the primary order parameter. The main distinction between the results of the impurity response of the two approaches is whether local anisotropies exist in the total LDOS or only in the spin-resolved LDOS. We find that the total LDOS exhibits direct local signatures of the altermagnetic order, whereas the one-band models feature such signatures only in the spin-resolved impurity response.

The paper is organized as follows. In Sec.~\ref{sec:methods} we introduce the basic methodology, including both the $T$-matrix formalism of a single impurity and the self-consistent real-space formulation of the problem. Section~\ref{sec:results} presents our main results for total and spin-resolved LDOS near point-like disorder sites in both $d$-wave and $g$-wave altermagnets. Altermagnetic symmetries are directly imprinted onto the LDOS near impurities. In Sec.~\ref{sec:singleband} and \ref{sec:QPI} we contrast the results to those obtained within a single-band model and in Sec.~\ref{sec:superconductivity} we discuss potential interplay with superconductivity. Finally, Sec.~\ref{sec:conclusions} contains our conclusions.

\section{Methodology}\label{sec:methods}
\begin{figure*}
    \centering
    \includegraphics[width=\linewidth]{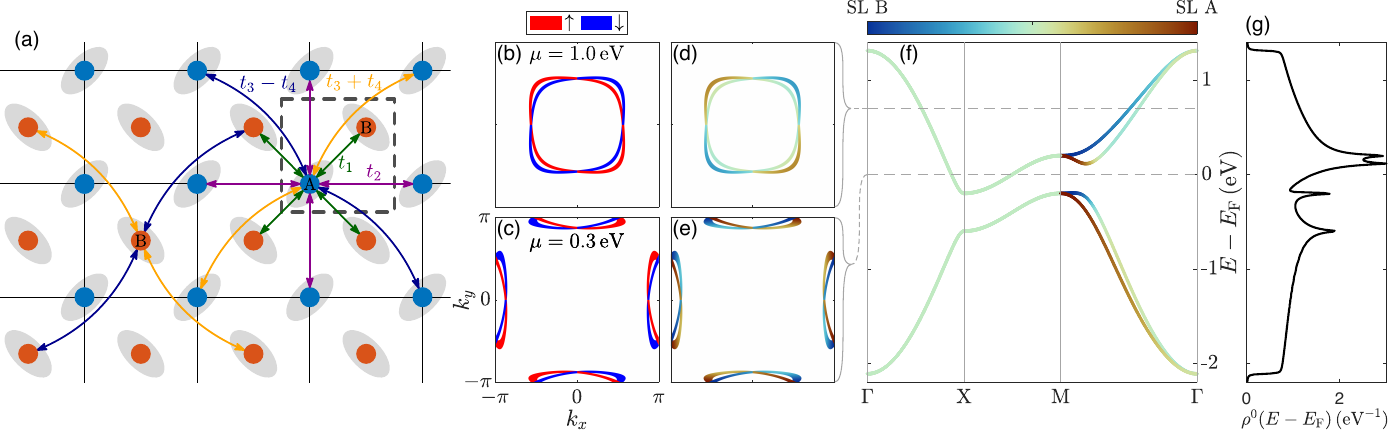}
    \caption{(a) Sketch of the crystal lattice and the hopping parameters representing layer group 63 with lattice sites at Wyckoff position 2b. The unit cell is indicated by the  dashed box. The hopping term responsible for altermagnetism $t_z \propto t_4$ has opposite signs for sublattices A and B and breaks $C_4$ symmetry~\cite{Roig_Minimal_2024}. The sublattices are related by translation combined with rotation by 90\textdegree{} due to the anisotropic nonmagnetic environment indicated by gray ellipses. (b), (c) Fermi surface for $\mu=1$\,eV and $\mu=0.3$\,eV colored by spin polarization. (d), (e) Same as (b), (c) but colored by sublattice weight. (f) Dispersion along high symmetry path colored by sublattice weight. (g) Density of states for the bands shown in (f).}
    \label{fig:model}
\end{figure*}
Throughout this study we begin with the minimal model for altermagnets derived in Ref.~\cite{Roig_Minimal_2024}
\begin{equation}
    \mathcal{H}_\mathrm{MM}=\varepsilon_{0,\mathbf{k}} \tau_0 +t_{x,\mathbf{k}} \tau_x + t_{z,\mathbf{k}} \tau_z  + \tau_y \vec{\lambda}_\mathbf{k}\cdot \vec{\sigma} + \tau_z \vec{N} \cdot \vec{\sigma},
\end{equation}
where $\sigma_i$ and $\tau_i$ are the Pauli matrices for the spin and sublattice degrees of freedom, respectively. The energy dispersion $\varepsilon_{0,\mathbf{k}}$ is independent of the sublattice and $t_{x,\mathbf{k}}$ and $t_{z,\mathbf{k}}$ denote inter- and intra-sublattice hoppings, respectively. The crystal asymmetric hopping $t_{z,\mathbf{k}}$ describes the local symmetry breaking already in the normal state, as it transforms as a non-trivial irreducible representation of the point group. Neglecting spin-orbit coupling given by the term $\vec{\lambda}_\mathbf{k}$, the quantization axis of the Néel order parameter $N$ can be chosen along the $z$ direction. To make Fourier transformations simpler and restrict $k$-space integrations to the first Brillouin zone (BZ) only, we perform a gauge transformation of the B sublattice creation and annihilation operators as $c_{\sigma B,\mathbf{k}}\rightarrow c_{\sigma B,\mathbf{k}} e^{i \phi_\mathbf{k}}$. In the new basis, $t_{x,\mathbf{k}}\rightarrow t_{x,\mathbf{k}}e^{i \phi_\mathbf{k}}$ such that the Hamiltonian exhibits the full periodicity of the BZ for $\phi_\mathbf{k}=-\frac{k_x}{2}-\frac{k_y}{2}$. The minimal model then reads
\begin{equation}\label{eq:minimalModel}
    \mathcal{H} = \varepsilon_{0,\mathbf{k}} \tau_0 + t_{z,\mathbf{k}} \tau_z + \Re[t_{x,\mathbf{k}}] \tau_x - \Im[t_{x,\mathbf{k}}] \tau_y + N \tau_z \sigma_z.
\end{equation}
The explicit momentum dependence of all expressions and tight-binding parameters are given in Appendix~\ref{TBparameter}. Note that $\mathcal{H}$ is block diagonal in the spin index $\sigma$ and each block can be diagonalized as $U_{\sigma, \mathbf{k}}^\dagger \mathcal{H}_\sigma U_{\sigma, \mathbf{k}} = \diag(E^{+}_{\sigma,\mathbf{k}},E^{-}_{\sigma,\mathbf{k}})$ with eigenvalues $E^\beta_{\sigma,\mathbf{k}}= \varepsilon_{0,\mathbf{k}} + \beta \sqrt{|t_{x,\mathbf{k}}|^2 + (t_{z,\mathbf{k}} + \sigma N)^2}$ and unitary transformation
\begin{equation}\label{eq:unitaryTrafo}
    U_{\sigma,\mathbf{k}} =
        \begin{pmatrix}
            \cos{\frac{\theta_\mathbf{\sigma,k}}{2}} & \frac{t_x}{\abs{t_x}}\sin{\frac{\theta_\mathbf{\sigma,k}}{2}} \\
            -\frac{t_x^*}{\abs{t_x}} \sin{\frac{\theta_\mathbf{\sigma,k}}{2}} & \cos{\frac{\theta_\mathbf{\sigma,k}}{2}}
        \end{pmatrix},
\end{equation}
where
\begin{align}
    \cos{\frac{\theta_\mathbf{\sigma,k}}{2}} &= \frac{1}{\sqrt{2}}\sqrt{1+\frac{t_z+\sigma N}{\sqrt{|t_x|^2 + (t_z + \sigma  N)^2}}},\\
    \sin{\frac{\theta_\mathbf{\sigma,k}}{2}} &= \frac{-1}{\sqrt{2}}\sqrt{1-\frac{t_z+\sigma N}{\sqrt{|t_x|^2 + (t_z + \sigma N)^2}}}, 
\end{align}
and $\beta \in \{-1,1\}$ is a band index.
Figure~\ref{fig:model} shows an exemplary $d$-wave altermagnet. The anisotropic crystal environments and hoppings for the sublattices are shown in Fig.~\ref{fig:model}(a). The spin-split Fermi surface and band structure can be seen in Fig.~\ref{fig:model}(b-f), also showing the sublattice mixing but pure spin character of the bands.

\subsection{T-matrix formalism}
To study local signatures of altermagnetic order close to impurities, we employ the $T$-matrix framework. We consider scattering with strength $V_{\mathrm{imp}}$ on a nonmagnetic onsite impurity potential in the elementary cell at the origin and at an A sublattice site with the Hamiltonian
$\mathcal{H}_\mathrm{imp}=\sum_iV_\mathrm{imp}\frac 12 (\tau_0+\tau_z) \delta_{i,0}\sigma_0$\footnote{In altermagnets, the sublattice sites are related by a proper (or improper) rotation in the normal state and by time-reversal and the corresponding proper (or improper) rotation. Thus, an impurity on a B sublattice exhibits the same response modulo the rotation operation in the normal state and modulo the rotation and time-reversal operation in the magnetic state.}.
The full Green's function in real space is given by
\begin{equation}\label{eq:GreensRealspace}
    \mathcal{G}(\mathbf{r},\mathbf{r}',\omega) = \mathcal{G}^0(\mathbf{r}-\mathbf{r}',\omega) + \mathcal{G}^0(\mathbf{r},\omega) T(\omega) \mathcal{G}^0(-\mathbf{r}',\omega),
\end{equation}
where $\mathbf{r}$ denotes unit cell positions and all quantities are matrices in spin- and sub-lattice space.
The $T$-matrix
\begin{equation}\label{eq:tmatrix}
    T(\omega) = \left[ 1 - \mathcal{H}_\mathrm{imp} \mathcal{G}^0(0,\omega) \right]^{-1} \mathcal{H}_\mathrm{imp},
\end{equation}
encompasses scatterings off the impurity. Both the bare Green's function $\mathcal{G}_\sigma ^0(\mathbf{r},\omega)$ and the scattering potential $\mathcal{H}_\mathrm{imp}$, and hence also the $T$-matrix, are diagonal in spin, so we can treat the equations for $\sigma = {\uparrow, \downarrow}$ independently.
The bare Green's function in sublattice space for spin $\sigma$ reads
\begin{equation}
    \bigl(\mathcal{G}_{\sigma}^0(\mathbf k)\bigr)_{ab}
    = \sum_\beta \frac{(U_{\sigma,\mathbf{k}})_{a \beta} (U_{\sigma,\mathbf{k}}^*)_{b\beta}}{\omega + i \eta -E^\beta_{\sigma,\mathbf{k}}}.
    \label{eq:GreensSL}
\end{equation}
The real-space expressions are obtained by Fourier transform, $\mathcal{G}_\sigma ^0(\mathbf{r},\omega)=\sum_{\mathbf k} e^{-i\mathbf k\cdot\mathbf r}\mathcal{G}_{\sigma}^0(\mathbf k)$.
This approach is applied to tight-binding models describing $d$-wave and $g$-wave altermagnets on a square lattice.
For the particular choice of $t_x$ used here, this applies to the following space groups and Wyckoff positions: SG 14 (2a-2d), SG 55 (2a-2d), SG 83 (2e-2f), SG 123 (2e-2f), and SG 127 (2a-2d). The associated band structure, Fermi surface (FS), and density of states (DOS) are summarized for this concrete example in Fig.~\ref{fig:model}. Parameters for the tight-binding model are given in Appendix~\ref{TBparameter}.

The spin-resolved LDOS at position $\mathbf{r}$ is obtained from $\rho_\sigma(\mathbf{r},\omega) = -\frac{1}{\pi} \Im{\mathcal{G}_\sigma(\mathbf{r},\mathbf{r},\omega)}$ and the homogeneous DOS far away from the impurity is given by $\rho^0_\sigma(\omega) = -\frac{1}{\pi} \Im{\mathcal{G}^0_\sigma(0,\omega)}$\footnote{Note that the position $\mathbf{r}$ refers to a position of a lattice point rather than an elementary cell and the corresponding matrix element of the Green function is meant on the r.h.s.}. We define the inhomogeneous part of the LDOS by subtracting the DOS far away from the impurity as $\delta\rho_\sigma(\mathbf{r},\omega) = \rho_\sigma(\mathbf{r},\omega) - \rho^0_\sigma(\omega)$.

To quantify the symmetry breaking in the LDOS, we further define
\begin{equation}
    \delta\rho_{\sigma,\mathrm{asym}}(\mathbf{r},\omega) = \delta\rho_\sigma(\mathbf{r},\omega)- \delta\rho_\sigma(\mathcal{S} \mathbf{r},\omega) ,
\end{equation}
where $\mathcal{S}$ is a symmetry operation that relates the altermagnetic hopping parameters of the two sublattices. For a $d$-wave altermagnet as shown in Fig.~\ref{fig:model}(a), this corresponds to a $C_4$ rotation or a mirror operation $M_x$ or $M_y$ along the lattice vectors. For a planar $g$-wave altermagnet, $\mathcal{S}$ is a mirror operation $M_x$, $M_y$ or along the diagonals $M_d$, $M_{d'}$. The spin-summed LDOS is $\rho=\sum_\sigma \rho_\sigma$ and all derived spin-summed quantities are calculated accordingly.

\subsection{Self-consistent real-space formulation}
A limitation of the $T$-matrix method is that it depends on the bare Green's function of the homogeneous lattice, and feedback effects due to the impurity potential close to the disorder site are not taken into account. To include effects from density modulations in the vicinity of the impurity at the mean-field level, we additionally solve the impurity problem starting from a real-space Hamiltonian $\mathcal{H}$ defined on an $N \times N$ grid with periodic boundary conditions using the same hopping integrals and parameters as before. Instead of fixing a global value for the order parameter $N$, we obtain it self-consistently by adding an on-site interaction term to the Hamiltonian $\mathcal{H} = \mathcal{H}_0 + \mathcal{H}_\mathrm{imp} + \mathcal{H}_\mathrm{int}$ for each lattice site individually~\cite{Roig_Minimal_2024}. The on-site Hubbard $U$ is added to the model and we perform a mean-field decoupling
\begin{align}
    \mathcal{H}_\mathrm{int} (\mathbf{r}_i) &= U n_{i \uparrow} n_{i \downarrow}\notag\\
    &\approx \sum_{\sigma} \frac{U}{2} \bigl( \expval{n_i} - \sigma N_i \bigr) n_{i \sigma} - U \expval{n_{i \downarrow}} \expval{n_{i \uparrow}},
\end{align}
where $\expval{n_i} = \expval{n_{i\uparrow} + n_{i\downarrow}}$ denotes the local electron filling and $N_i = \expval{n_{i\uparrow} - n_{i\downarrow}}$ the order parameter at site $i$. We neglect the constant term $U \expval{n_{i \downarrow}} \expval{n_{i \uparrow}}$ and absorb the term $\frac{U}{2} \expval{n_0}$ containing the average filling $\expval{n_0}$ into the chemical potential and keep $\expval{\delta n_i} = \expval{n_i} - \expval{n_0}$ for the self-consistency equation, so the interacting part of the Hamiltonian becomes in mean-field approximation
\begin{equation}
    \mathcal{H}_\mathrm{MF, int} \mathbf{r}_i) = \sum_{\sigma} \frac{U}{2} \bigl(\expval{\delta n_i} - \sigma N_i \bigr) n_{i \sigma}.
\end{equation}
We solve the resulting Hamiltonian $\mathcal{H} = \mathcal{H}_0 + \mathcal{H}_\mathrm{imp} + \mathcal{H}_\mathrm{MF, int}$ by numerical diagonalization, use the eigenvalues and eigenvectors to calculate the mean fields $\expval{n_i}$ and $N_i$  as thermal averages at temperature $T = 0.01$\,eV at fixed filling, i.\,e. $\sum_i \expval{\delta n_i}=0$  and
iterate until convergence. Afterwards, we use the eigenvalues and eigenvectors again to obtain the spatially resolved Green's function and the associated LDOS.

\section{Results}\label{sec:results}
\begin{figure}
    \centering
    \includegraphics[width=\linewidth]{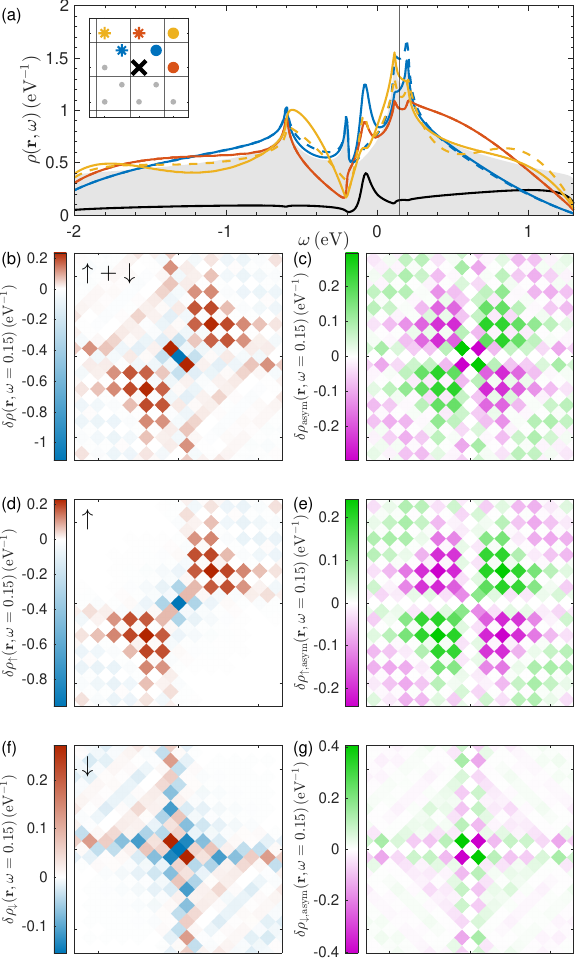}
    \caption{(a) LDOS as a function of energy in the vicinity of a nonmagnetic impurity with $V_\mathrm{imp}=1.8$\,eV. The colored circles (stars) in the inset indicate the site of the corresponding solid (dashed) line. Same colored sites are related by 90\textdegree{} rotation around the impurity. The homogeneous DOS far away from the impurity is shown in gray and the line at $\omega=0.15$\,eV indicates the energy chosen in panels (b)-(g). (b), (d) and (f) LDOS without homogeneous background in real-space summed over both spins, and for spin-up and spin-down electrons, respectively. (c), (e) and (g) difference of the LDOS and 90\textdegree{} rotated LDOS. Calculations are done with a grid size of $512\times 512$ $k$-points at $\eta=5$\,meV.}
    \label{fig:signaturesAM}
\end{figure}
\subsection{Signatures from altermagnetic minimal model}
We first apply the $T$-matrix formalism described above to a minimal model example featuring $d$-wave 
spin-splitting $t_z= 4 t_4 \sin k_x \sin k_y$.
In Fig.~\ref{fig:signaturesAM}(a) we show the calculated
LDOS at representative sites close to an impurity with repulsive potential of $V_\mathrm{imp}=1.8$\,eV.
The nonmagnetic impurity induces a symmetry breaking of the LDOS close to the disorder site as seen by the splitting of the results on the sites of same color (inset of Fig.~\ref{fig:signaturesAM}(a)) which are connected by a 90 degrees rotation (full line, dashed). This signature is present at all energies and is not qualitatively affected by the choice of impurity potential. Far away from the impurity, the DOS will converge to the bulk DOS shown in Fig.~\ref{fig:model}(g). For further analysis and presentation of LDOS maps, we choose a probe energy of $\omega=0.15$\,eV as there the symmetry breaking is rather large. However, the symmetry statements are valid across all energies. Importantly, breaking of $C_4$ symmetry is not only present in the spin-resolved LDOS evident from Fig.~\ref{fig:signaturesAM}(d)-(g), but also in the total LDOS as seen from Fig.~\ref{fig:signaturesAM}(b),(c). As seen from Figs.~\ref{fig:signaturesAM}(c),(e) and (f) the symmetry and the nodal lines of the $d$-wave altermagnet are directly imprinted in the total LDOS where $\delta\rho_{\mathrm{asym}}$ vanishes. In this respect, local nonmagnetic disorder acts as local signature of the underlying altermagnetic order. For disorder located on the B sublattice sites, the resulting LDOS is identical to those presented in Fig.~\ref{fig:signaturesAM} upon a $C_4$ rotation and exchange of spin component.
$d$-wave altermagnets with $t_z=t_4(\cos{k_x} -\cos{k_y})$ spin-splitting can be obtained from models classified by layer groups L61 and a special case of L51 (see Tab.~\ref{tab:layerGroups}), but are not further discussed here. In this case, the imprinted nodal lines will be 45\textdegree{} rotated compared to the $d$-wave case discussed above.

\begin{figure}
    \centering
    \includegraphics[width=\linewidth]{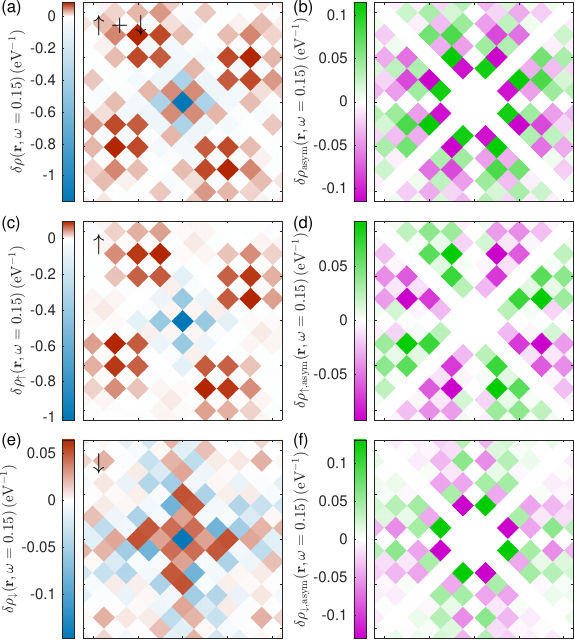}
    \caption{(a), (c) and (e) LDOS without homogeneous background for an altermagnet with $g$-wave splitting spin-summed and spin resolved for spin-up and spin-down, respectively.
    (b) Difference of the LDOS and LDOS mirrored along one of the $g$-wave nodal lines of the left column figures.}
    \label{fig:signaturesAM-gwave}
\end{figure}
The resulting LDOS impurity effect is also calculated for a $g$-wave altermagnet where the parameters from Tab.~\ref{tab:parameters} are used together with $t_z=t_4 \sin k_x \sin k_y (\cos k_x-\cos k_y)$.
Fig.~\ref{fig:signaturesAM-gwave}(a) shows the DOS modulations in this $g$-wave altermagnet.
From $\delta\rho_\mathrm{asym}$ as seen in Fig.~\ref{fig:signaturesAM-gwave}(b) it is evident that there is no effect along the horizontal, vertical and diagonal directions which imprints the nodal lines of the $g$-wave spin-splitting.
The spin-up and spin-down DOS are not equally affected by the impurity and their contributions to the total LDOS can interfere constructively or destructively for different sites, meaning the pattern of the $\delta\rho_\sigma$ is different, but they break the same symmetry. For the particular choice of $\omega$ and $V_{\mathrm{imp}}$, the breaking of the LDOS is strongest for spin down (see Figs.~\ref{fig:signaturesAM}(g) and \ref{fig:signaturesAM-gwave}(f)).\par
In all cases, nonmagnetic impurities uncover not only the presence of altermagnetism but also indicate the underlying altermagnetic order itself.
The LDOS pattern for an impurity on a B or A sublattice is related by the same symmetry operation as the sublattices A and B themselves, i.\,e. an exchange of the spins combined with a $C_4$ rotation for the $d$-wave case and a mirror symmetry for the $g$-wave case.

\subsection{Altermagnet with vanishing order parameter}
\begin{figure}
    \centering
    \includegraphics[width=\linewidth]{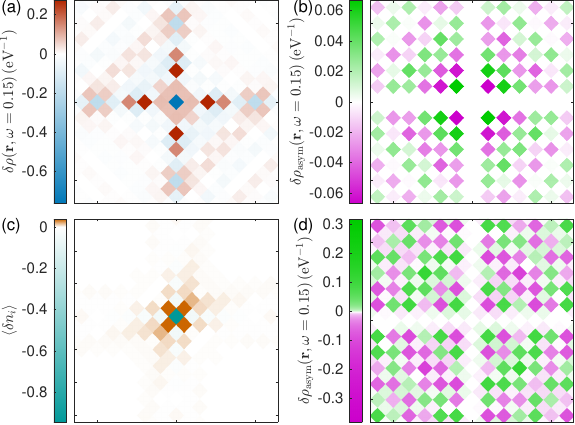}
    \caption{(a) LDOS from a $T$-matrix calculation without homogeneous background for an altermagnet in the NS (i.\,e. $N=0$) in real space and (b) LDOS asymmetry with a signal only on the A sublattice. (c) Deviation of electron density from self-consistent mean-field calculation with $64\times 64$ elementary cells and $U=0.9$\,eV and (d) resulting asymmetric part of the LDOS. The colorbar in (d) is logarithmic to accentuate the small values on the B sublattice.}
    \label{fig:signaturesAM-NS}
\end{figure}

Is the impurity-induced symmetry breaking exclusive to the magnetic state or can it also manifest in the absence of magnetism? To examine this question, we study the case of vanishing order parameter $N=0$ and keep the $t_z$ term in the tight-binding model since it is generated by the local environments of the atoms of the sublattice. Note that the eigenenergies will be spin degenerate in this case and there are no spin split bands. Yet, the LDOS in Fig.~\ref{fig:signaturesAM-NS}(a) shows again breaking of $C_4$ symmetry but $\delta{\rho}_\mathrm{asym}$ is finite only on the sites of the same sublattice as the impurity and zero on all sites of the opposite sublattice sites, apparent in the checkerboard pattern in Fig.~\ref{fig:signaturesAM-NS}(b).

We sketch an explanation as follows: Each scattering process of an electron on a B sublattice site with the impurity on the A sublattice must involve an odd number of inter-sublattice hopping processes $t_{A\leftrightarrow B}$, given by the $t_x$ term. Scattering processes including $t_z$ can potentially be symmetry breaking. While $t_z$ commutes with intra-sublattice hoppings, we note that they anti-commute with inter-sublattice hoppings, meaning $t_{A\leftrightarrow B} t_z = -t_z t_{A\leftrightarrow B}$. When summing over all possible scattering processes, this fact together with the odd number of sublattice changes makes all terms containing $t_z$ cancel and hence the LDOS on the B sublattice does not break any of the point group symmetries. For scattering processes between the impurity and an A sublattice site there is an even number of hopping processes that change the sublattice, and we observe symmetry breaking as shown in Fig.~\ref{fig:signaturesAM-NS}(b). This argument is based on the fact that we assume hoppings to be homogeneous regardless of the impurity site and the $T$-matrix formalism considers the homogeneous, bare Green's function. In reality, the presence of an impurity strongly affects hopping processes in its vicinity.

To take this into account and study its effect, we consider the system in real space and calculate the order parameter self-consistently. Treating the order parameter on a mean-field level gives a magnetic transition at $U \approx 0.95$\,eV. For $U=0.9$\,eV the system is in the nonmagnetic normal state and the self-consistent order parameter is numerically zero, yet the electron density shifts close to the impurity and breaks $C_4$ symmetry as shown in Fig.~\ref{fig:signaturesAM-NS}(c). Calculating the Green's function from this real space Hamiltonian via exact diagonalization now reveals symmetry breaking of the LDOS on both sublattices, but the effect is still more than one order of magnitude smaller on the B sublattice, see Fig.~\ref{fig:signaturesAM-NS}(d).

Consequently, a symmetry-broken LDOS close to an impurity site is not sufficient to indicate altermagnetism, as it can also occur in the nonmagnetic state. Additionally, Néel order must be shown by experimental techniques like neutron scattering to have conclusive evidence of altermagnetism. In addition, even in the absence of magnetism, symmetry breaking of the LDOS indicates the presence and strength of the altermagnetic hopping $t_z$ which induces the spin splitting once the material becomes magnetic at lower temperatures or by doping.

In the normal state close to the transition, magnetic disorder can induce altermagnetic order locally. For some systems, it is well known that nonmagnetic impurities can also pin local regions of the low-temperature ordered phase~\cite{Chen2004,Harter2007,Andersen2007,Andersen2008,Schmid_2010,Martiny2015,Gastiasoro2016,Martiny2019}. For the present model for altermagnetism, however, we have not found evidence for this effect. 

\subsection{Impurity response from one-band model}\label{sec:singleband}
\begin{figure}
    \centering
    \includegraphics[width=\linewidth]{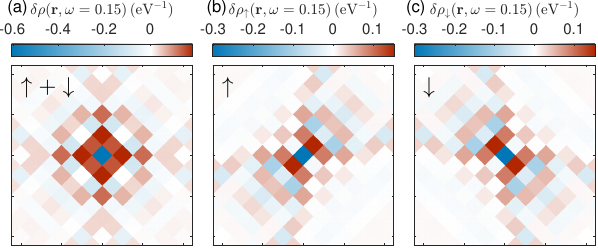}
    \caption{(a) Spin-summed and (b,c) spin-resolved LDOS for a nonmagnetic impurity in the one-band model, Eq.~(\ref{oneband}). The symmetry breaking occurs in each of the spin channels separately, but cancels in the total spin-summed LDOS in contrast to models built on the sublattice degrees of freedom, see Figs.~\ref{fig:signaturesAM}-\ref{fig:signaturesAM-gwave}.}
    \label{fig:signaturesAM-1SL}
\end{figure}

A frequently used model for altermagnetic metals in the literature reduces complexity by applying a one-band model formulated in band space. Instead of exhibiting a sublattice degree of freedom, it is assumed that only a single band crosses the Fermi level and that it exhibits a momentum-dependent spin splitting as recently classified and expressed in an effective momentum-dependent magnetic field~\cite{Radaelli2024}. In this case the Hamiltonian is of the form
\begin{equation}\label{oneband}
    \mathcal{H}_\mathrm{1SL} = \varepsilon_{0,\mathbf{k}} \sigma_0 + N t_{z,\mathbf{k}}\sigma_z .
\end{equation}
The specific terms and tight-binding parameters are listed in Appendix~\ref{TBparameter}. In this subsection we study the local impurity-induced signatures obtained within such a one-band model and compare them to the results obtained within the minimal models based on sublattice site-symmetries. Due to the lack of sublattice degree of freedom, the form of the potential scatterer for this model is $\mathcal{H}_\mathrm{imp}=\sum_i V_\mathrm{imp} \delta_{i,0}\sigma_0$.
The main distinction between the two models can be inferred from Fig.~\ref{fig:signaturesAM-1SL}. For an example case of $d$-wave spin-splitting, it is observed that the symmetry is broken only in the separate spin channels, as seen from Fig.~\ref{fig:signaturesAM-1SL}(b,c), and that they cancel in the spin-summed LDOS as evident from Fig.~\ref{fig:signaturesAM-1SL}(a). This highlights a fundamental difference compared to results obtained from models based on the underlying sublattice degree of freedom. In addition, in the normal state, i.\,e. without Néel order and spin-split bands, there cannot be a symmetry breaking from the impurity within the one-band model, in contrast to the results of Fig.~\ref{fig:signaturesAM-NS}.

\subsection{Quasiparticle interference}\label{sec:QPI}
The same arguments hold for quasi-particle interference (QPI) in altermagnets: while one-band models produce anisotropy only in the spin-split LDOS response~\cite{Hu_Quasiparticle_2025,Chen_impurity2024}, models that are based on the sublattice degrees of freedom will feature spatial anisotropy already in the total LDOS. Intuitively, this can be seen directly from Fig.~\ref{fig:model}(d,e) highlighting the sublattice character of the states at the Fermi level.

\begin{figure}
    \centering
    \includegraphics[width=\linewidth]{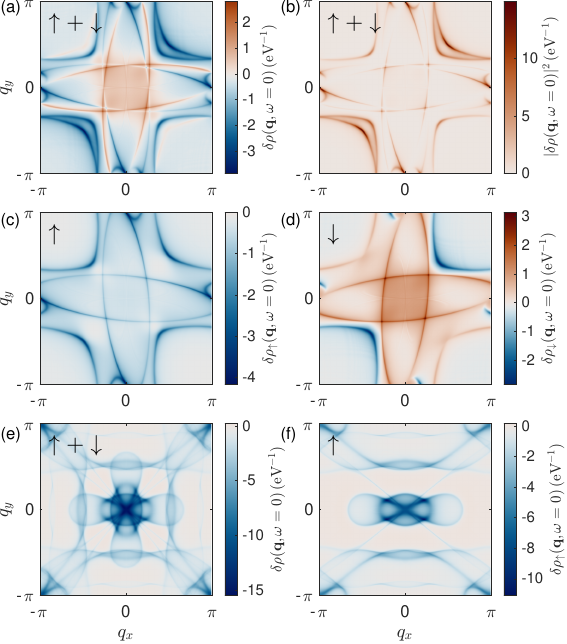}
    \caption{(a) Spin-summed QPI spectrum for an impurity in a two-band $d$-wave altermagnet. (b) Power spectrum of panel (a) as measured in experiments.
    (c,d) spin-resolved QPI spectra of panel (a).
    (e) Spin-summed QPI spectrum in the single-band model and (f) spin-resolved QPI spectum showing only the spin up contribution in panel (e). We use $301\times 301$ unit cells for the calculation of the spectra.}
    \label{fig:QPI}
\end{figure}

To sustain this argument, we calculate the QPI spectra by Fourier transforming the LDOS and omit the translation invariant term in Eq.~\eqref{eq:GreensRealspace}, as it only contributes to a Bragg peak at $\mathbf{q}=0$. The density modulations in momentum space read
\begin{equation}
    \delta\rho(\mathbf{q},\omega)=-\frac{1}{\pi} \sum_\mathbf{r} e^{-i \mathbf{q} \cdot \mathbf{r}}\Im[\mathcal{G}^0(\mathbf{r},\omega) T(\omega) \mathcal{G}^0(-\mathbf{r},\omega)].
\end{equation}
We present the QPI spectra at $\omega=0$ of a $d$-wave altermagnet for the model including a sublattice degree of freedom (Eq.~\eqref{eq:minimalModel}) and the single-band model (Eq.~\eqref{oneband}) in Fig.~\ref{fig:QPI}.
Again, a symmetry breaking can be observed already in the spin-summed spectrum for the two-band model as seen in Fig.~\ref{fig:QPI}(a-d) whereas for the single-band model the symmetry is broken only in the spin-resolved spectrum as shown in Figs.~\ref{fig:QPI}(e,f), since the spectra for the spin up and down channels are related by a 90\textdegree{} rotation. For better comparability to scanning tunneling conductance experiments, we also show the calculated power spectrum as the amplitude squared of the QPI signal in Fig.~\ref{fig:QPI}(b) for the two-band model. Furthermore, the spectra for impurities on opposite sublattices are symmetry related (here a $C_4$ rotation of Figs.~\ref{fig:QPI}(a,b)). We point out that symmetry breaking effects average out for large scanning areas for similar distribution of impurities on the sublattices.
To detect the symmetry breaking, it might be needed to examine small fields of view with imbalanced number of impurities on the two sublattices, or experimentally assign the impurities to the sublattices and selectively perform the Fourier transformation as presented in Ref.~\cite{Chen2023}.

\subsection{Interplay with superconductivity}\label{sec:superconductivity}
\begin{figure}
    \centering
    \includegraphics[width=\linewidth]{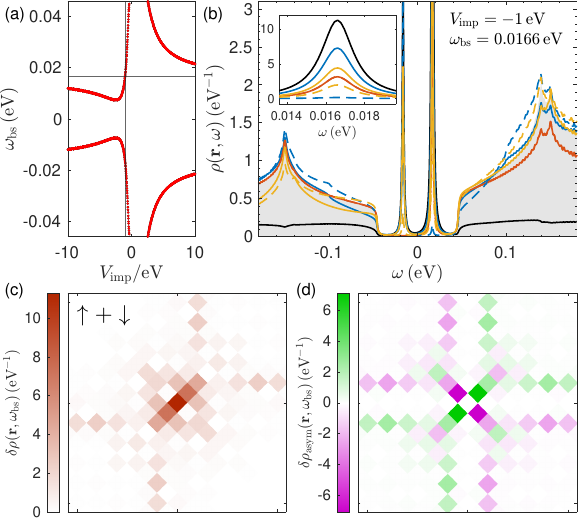}
    \caption{(a) Bound state energy as function of impurity potential, the vertical and horizontal line indicate the impurity potential and corresponding bound state energy used for the other panels, (b) LDOS at lattice sites close to the impurity, same colored sites are related by 90\textdegree{} rotation with same color code as in \ref{fig:signaturesAM}(a). The inset shows magnification near the bound state energy $\omega_\mathrm{bs}=16.6$\,meV at $V_\mathrm{imp}=-1$\,eV both indicated by lines in panel (a). (c) and (d) LDOS and asymmetric part of the LDOS at $\omega_\mathrm{bs}$, respectively. Calculations are done with a grid size of $1024\times 1024$ $k$-points at $\eta=1$\,meV.}
    \label{fig:signaturesAM-SC}
\end{figure}

The effect of impurities in gapped systems can lead to very sharp impurity bound states as exemplified by sharp impurity resonances at low energies in superconductors, which can be used to amplify signatures of underlying ordered states~\cite{GastiasoroFK,Gastiasoro2017,Chen2023}.
We therefore incorporate superconductivity from an altermagnetic metallic phase~\cite{Brekke_Twodimensional_2023}, e.\,g. by proximity effect or intrinsic Cooper pairing instability by use of the Nambu formalism. We aim to include only spin-singlet superconductivity, allowing us to decompose the Bogoliubov–de Gennes (BdG) Hamiltonian with sublattice, spin and Nambu degrees of freedom into two $4 \times 4$ blocks
\begin{equation} \label{eq:BdG}
    \mathcal{H}_\mathrm{BdG}=
    \begin{pmatrix}
        \mathcal{H}_{\uparrow,\mathbf{k}} & \Delta_\mathbf{k} \\
        \Delta^\dagger_\mathbf{k} & -\mathcal{H}^T_{\downarrow,-\mathbf{k}}
    \end{pmatrix},
\end{equation}
in the Nambu spinor basis $\Psi^\dagger_\mathbf{k}=(c^{A\dagger}_{\uparrow \mathbf{k}}, c^{B\dagger}_{\uparrow \mathbf{k}}, c^{A}_{\downarrow -\mathbf{k}}, c^{B}_{\downarrow -\mathbf{k}})$.
Here, we write the BdG Hamiltonian for spin up electrons and spin down holes. The second, equivalent block of the BdG Hamiltonian is obtained by reversing the spins. For the SC gap, we consider a simple constant order parameter in band space $\Delta_\mathrm{b}=\Delta\mathbb{1}$.
Using the unitary transformation Eq. \eqref{eq:unitaryTrafo} of the normal state, we back transform $\Delta_\mathrm{b}$ and obtain the corresponding order parameter $\Delta_\mathbf{k}$ in sublattice space
\begin{equation}\label{eq:gapsublattice}
    \Delta_\mathbf{k}=U_{\uparrow,\mathbf{k}} \Delta_\mathrm{b} U_{\downarrow,-\mathbf{k}}^T,
\end{equation}
which becomes momentum dependent and can be used for input to Eq.~(\ref{eq:BdG}).
The band gap edge and the coherence peak shift approximately linear with the maximum of $t_z$ at the Fermi level. When the altermagnetic order parameter becomes large, the gap eventually closes and the superconducting state features a Bogoliubov Fermi surface~\cite{Hong2025} as generically expected in superconductors with broken time-reversal symmetry~\cite{Agterberg:2017,Setty2020}. Using the expression Eq.~(\ref{eq:gapsublattice}) within the $T$-matrix approach, in-gap bound states are found where the real part of the $T$-matrix in Eq.~(\ref{eq:tmatrix}) exhibits poles.
The nonmagnetic impurity potential in the Nambu spinor basis is $\mathcal{H}_\mathrm{imp}=V_\mathrm{imp}\frac 12 (\tau_0+\tau_z) \sigma_z$ \footnote{Note that here $\sigma_z$ refers to particle-hole space. By choice of the basis for Eq.~\eqref{eq:BdG}, electrons carry spin up and holes carry spin down.}.
We solve for the real roots of the $T$-matrix to calculate the bound state energy $\omega_{\mathrm{bs}}$ for a range of impurity potentials at a chemical potential $\mu=0.3$\,eV.
Examining the result in Fig.~\ref{fig:signaturesAM-SC}(a), one sees that the altermagnetic order allows for the existence of in-gap bound states despite the nonmagnetic nature of the impurities~\cite{Maiani}. The corresponding LDOS as calculated from the $T$-matrix approach exhibits pairs of sharp peaks at $\pm\omega_{\mathrm{bs}}$ in the gapped region, see  Fig.~\ref{fig:signaturesAM-SC}(b). We note, however, that the in-gap bound states are not guaranteed by symmetry and depend on the dispersion and chemical potential. In the nonmagnetic state, the transformation in Eq.~(\ref{eq:gapsublattice}) becomes trivial with $\Delta_\mathbf{k}=\Delta_\mathrm{b}$ and no bound states are observed. At the bound state energies, the disorder-induced symmetry breaking of the LDOS is strongly enhanced as evident from Fig.~\ref{fig:signaturesAM-SC}(c,d), making proximity-induced superconductivity in altermagnets an interesting platform to detect local signatures of altermagnetism.

\section{Conclusions}\label{sec:conclusions}
We have used microscopic models for altermagnetic materials based on the relationship between the site symmetry of magnetic atoms and the point group symmetry of the associated space group, in order to study their real-space fingerprints in terms of the modified local density of states near impurity sites. We found the momentum structure of the spin-dependent band splitting to be directly imprinted on the total LDOS around the impurity site. This finding suggests that scanning tunneling conductance experiments can be used to detect altermagnets, a property that may be particularly useful in materials where domains average out the generic properties of altermagnets. Furthermore, we demonstrate that the lower site symmetry is imprinted on impurity states already without magnetic order whereas screening from the intinerant electrons make this signal very weak. Gapping of the electronic structure allows for sharp impurity bound states such that already conventional superconductivity can easily induce low-energy bound states from potential scatterers.

\begin{acknowledgments}
J.G. acknowledges support from the Independent Research Fund Denmark, Grant No. 3103-00008B. A.K. acknowledges support by the Danish National Committee for Research Infrastructure (NUFI) through the ESS-Lighthouse Q-MAT. 
D.F.A was supported by National Science Foundation Grant No. DMREF 2323857. Work at UWM  was also supported by a grant from the Simons Foundation (SFI-MPS-NFS-00006741-02, M.R. and Y.Y.).
\end{acknowledgments}

\appendix
\renewcommand{\thefigure}{S\arabic{figure}}
\setcounter{figure}{0}

\section{Tight-binding parameters} \label{TBparameter}
The terms in the tight binding Hamiltonian \eqref{eq:minimalModel} are given by
\begin{align}
    \varepsilon_0 &= -2 t_2(\cos{k_x}+\cos{k_y})-4 t_3 \cos{k_x}\cos{k_y} -\mu,\\
    \begin{split}
         t_x&= -t_1 \left(1+e^{-ik_x}+e^{-ik_y}+e^{-i(k_x+k_y)} \right) \\
         &= -4t_1 e^{-i(\frac{k_x}{2}+\frac{k_y}{2})} \cos{\frac{k_x}{2}}\cos{\frac{k_y}{2}},
    \end{split}\\
    t_z &= 4 t_4 \sin{k_x} \sin{k_y},
\end{align}
as illustrated in Fig.~\ref{fig:model}. The values of the hopping parameters are given in Tab.~\ref{tab:parameters} and chosen the same as in the 2D model from Ref.~\cite{Roig_Minimal_2024}.\par

For the single band Hamiltonian in Eq.~\eqref{oneband}, the terms are given by
\begin{align}
    \varepsilon_0 &=
    \begin{aligned}[t]
        &-2 t_1(\cos{k_x}+\cos{k_y})-4t_2 \cos{k_x}\cos{k_y}\\
        &-2t_3 (\cos{2 k_x}+\cos{2k_y}) -\mu,
    \end{aligned}\\
    N t_z &= 4 t_4 \sin{k_x} \sin{k_y},
\end{align}
with parameters specified in Tab.~\ref{tab:parameters}. Note that the BZ is different from the one of Eq.~\eqref{eq:minimalModel} since there is only one site per unit cell.
\begin{table}[h]
\caption{Hopping parameters for the minimal model [see Eq.~(\ref{eq:minimalModel})] in eV.}
\label{tab:parameters}
\begin{tabularx}{0.9\linewidth}{*{6}{>{\centering\arraybackslash} X}}
    \toprule
    $t_1$ & $t_2$ & $t_3$ & $t_4$ & $\mu$ & $N$  \\
     $0.425$ & $0.05$ & $-0.025$ & $-0.075$ & $0.3$ & $0.2$\\
     \bottomrule
\end{tabularx}
\end{table}

\section{Two-dimensional layer groups}\label{Yuetable}

For numerical calculations, 2D minimal models are useful. Using our earlier approach for developing minimal models~\cite{Roig_Minimal_2024},  we have developed 2D layer group-based minimal models. These models are written in Tab.~\ref{tab:layerGroups} for all 2D layer groups that have a primitive lattice and also have Wyckoff positions of order 2 that contain inversion in the Wyckoff site symmetry group.
\newcolumntype{Y}{>{\centering\arraybackslash}X}
\begin{table}[h]
\centering
	\caption{Tight-binding coefficients for 2D layer groups and Wyckoff positions with two atoms per unit cell at the inversion center. Abbreviation $c_i\equiv\cos k_i$, $s_i\equiv\sin k_i$, $c_{i/2}\equiv\cos \frac{k_i}{2}$, $s_{i/2}\equiv\sin \frac{k_i}{2}$ applies. Altermagnets with $d$-wave spin-splitting can be found in L15, L16, L17, L40, L44, L51, L61, and L63, while $g$-wave altermagnets can be found in L63. Coefficients are omitted for simplicity. We have included entries which are 0 in the  $\tau_z$ column. These entries have magnetic ground states that can be classified as altermagnetic, but exhibit no spin-splitting.}
    \label{tab:layerGroups}
\begin{tabularx}{\linewidth}{c c c Y}
\toprule
\makecell{2D layer\\groups}& \makecell{Wyckoff\\ positions}&$\tau_x$&$\tau_z$ \\ \midrule
L7 (p112/a)& 2a-2b&$c_{x/2}$&$0$\\ 

L15 (p2$_1$/m11)& 2a-2b&$c_{x/2}$&$s_xs_y$\\ 

L16 (p2/b11)& 2a-2b&$c_{y/2}$&$s_xs_y$\\ 

L17 (p2$_1$/b11) & 2a-2b&$c_{x/2}c_{y/2}$&$s_xs_y$\\ 

L38 (pmaa)& 2a-2b&$c_{x/2}$&$0$\\ 

L40 (pmam)& 2a-2b&$c_{x/2}$&$s_xs_y$\\ 

L41 (pmma)& 2a-2b&$c_{x/2}$&0\\ 

L42 (pman)& 2a-2b&$c_{x/2}c_{y/2}$&0\\ 

L44 (pbam)& 2a-2b&$c_{x/2}c_{y/2}$&$s_xs_y$\\ 

L51 (p4/m)& 2c&$c_{x/2}c_{y/2}$&$t_{z1}(c_x-c_y) +t_{z2}s_xs_y$\\ 
 
L61 (p4/mmm)& 2c&$c_{x/2}c_{y/2}$&$c_x-c_y$\\ 

L63 (p4/mbm)& 2a&$c_{x/2}c_{y/2}$&$s_xs_y(c_x-c_y)$\\ 

L63 (p4/mbm)& 2b&$c_{x/2}c_{y/2}$&$s_xs_y$\\ \bottomrule
\end{tabularx}
\end{table}

\bibliography{references}
\end{document}